# Photonics-Assisted Joint Communication-Radar System Based on a QPSK-Sliced Linearly Frequency-Modulated Signal

Shi Wang[a,b], Dingding Liang[a,b], and Yang Chen[a,b,*]

[a] *Shanghai Key Laboratory of Multidimensional Information Processing, East China Normal University, Shanghai 200241, China*
[b] *Engineering Center of SHMEC for Space Information and GNSS, East China Normal University, Shanghai 200241, China*
[*] *ychen@ce.ecnu.edu.cn*

**ABSTRACT**
A photonics-assisted joint communication-radar system is proposed and experimentally demonstrated, by introducing a quadrature phase-shift keying (QPSK)-sliced linearly frequency-modulated (LFM) signal. An LFM signal is carrier-suppressed single-sideband modulated onto the optical carrier in one dual-parallel Mach-Zehnder modulator (DPMZM) of a dual-polarization dual-parallel Mach-Zehnder modulator (DPol-DPMZM). The other DPMZM of the DPol-DPMZM is biased as an IQ modulator to implement QPSK modulation on the optical carrier. The polarization orthogonal optical signals from the DPol-DPMZM are further combined and detected in a photodetector to generate the QPSK-sliced LFM signal, which is used to realize efficient data transmission and high-performance radar functions including ranging and imaging. An experiment is carried out. Radar range detection with an error of less than 4 cm, ISAR imaging with a resolution of 14.99 cm × 3.25 cm, and communication with a data rate of 105.26 Mbit/s are successfully verified.

## 1. Introduction

With the rapid development of electronic technology, the combat environment of electronic warfare is becoming more and more complex. Faced with increasing threats and a complex electromagnetic environment, combat platforms have to be equipped with more and more electronic equipment, such as radar systems and communications systems, which brings serious electromagnetic compatibility problems. Therefore, it is of great significance to integrate multiple functions into a single system. Both radar systems and communication systems achieve their functions by transmitting and receiving electromagnetic waves. Therefore, the joint communication-radar system allows the communication system and the radar system to share the transmitting antenna, as well as the transmitter, which improves the system integration [1], [2] and reduces the system costs. The reported joint communication-radar system can be divided into two categories: signal multiplexing [3], [4] and signal sharing [5]-[8]. The former achieves a dual-functional system by using different multiplexing techniques. However, the radar signal and communication signal are generated independently, which do not take full advantage of the hardware of the system. The latter is based on a dual-functional integrated signal, which can be used for both radar detection and information transmission. Nevertheless, the conventional electrical signal generation methods suffer from electronic bottlenecks and are susceptible to the limitations of the operating frequency band and bandwidth.

Microwave photonics can effectively overcome the electronic bottleneck faced by conventional electronic systems due to its unique advantages in generating, transmitting, and processing microwave signals [9]-[12]. Moreover, the development of millimeter-wave communication technology [13], [14] and microwave photonic radar [15], [16] push the boundaries of the joint communication-radar system. In the past few years, the dual-functional waveform generation for the joint communication-radar system with the assistance of microwave photonics for the joint communication-radar system has been studied in [17]-[19]. Data information for the communication system can be inserted into a linearly frequency-modulated (LFM) signal for the radar system by controlling its amplitude or phase [17], [18]. We can also directly use the phase-coded signal, that is, the phase-shift keying signal of the communication system to realize the functions of radar and communication at the same time [19]. Based on the study of joint communication-radar waveform generation, joint communication-radar systems have been demonstrated in recent years [20]-[24]. Radar imaging and wireless communication were simultaneously achieved by modulating the binary data for communication onto a continuous-wave linearly frequency-modulated radar signal [20]. However, due to the employment of amplitude-shift keying modulation, the efficiency of data transmission is limited. Orthogonal frequency division multiplexing (OFDM) signals allowed not only data communication but also radar detection by extracting target information through the relative phase shifts of different subcarriers in [21]. In [22], a joint communication-radar system based on an optoelectronic oscillator (OEO) was proposed, in which quadrature phase-shift keying (QPSK) signals were used to enable distance measurement and communication functions. However, the radar imaging of the joint communication-radar system was not verified in [22] because the application of QPSK signals in radar imaging is more complex than that of conventionally used LFM signals. In [23], a joint communication-radar system was proposed, in which OFDM signals and LFM signals were integrated by time-division multiplexing (TDM) and used as the transmitted signal to realize the distance measurement and wireless communication. Indeed, the transmitted signal in [23] is obtained by splicing two kinds of signals using TDM instead of generating an integrated signal. In [24], photonic phase-coding and spectrum-spreading multiplexing techniques were incorporated to generate the joint communication-radar signals to enable radar and communication functions. Same as that in [22], radar imaging was not demonstrated in [24]. Therefore, it is urgent to study joint communication-radar systems which can simultaneously realize efficient data transmission and high-performance radar functions including ranging and imaging.

In this paper, we propose and experimentally demonstrate a joint communication-radar system using a QPSK-sliced LFM signal for radar target detection and wireless communications. An LFM signal is carrier-suppressed single-sideband (CS-SSB) modulated onto the optical carrier in one dual-parallel Mach-Zehnder modulator (DPMZM) of a dual-polarization dual-parallel Mach-Zehnder modulator (DPol-DPMZM). The other DPMZM of the DPol-DPMZM is biased as an IQ modulator to implement QPSK modulation on the optical carrier. The polarization orthogonal optical signals from the DPol-DPMZM are further combined and detected in a photodetector (PD) to generate the QPSK-sliced LFM signal, which is used for both wireless data transmission and radar detection. An experiment is carried out. Radar range detection and inverse synthetic aperture radar (ISAR) imaging are implemented with an accuracy of 4 cm and an imaging resolution of 14.99 cm×3.25 cm. The wireless communication capability is also verified by transmitting and receiving a 105.26-Mbit/s QPSK-sliced LFM signal in the wireless channel.

## 2. Principle

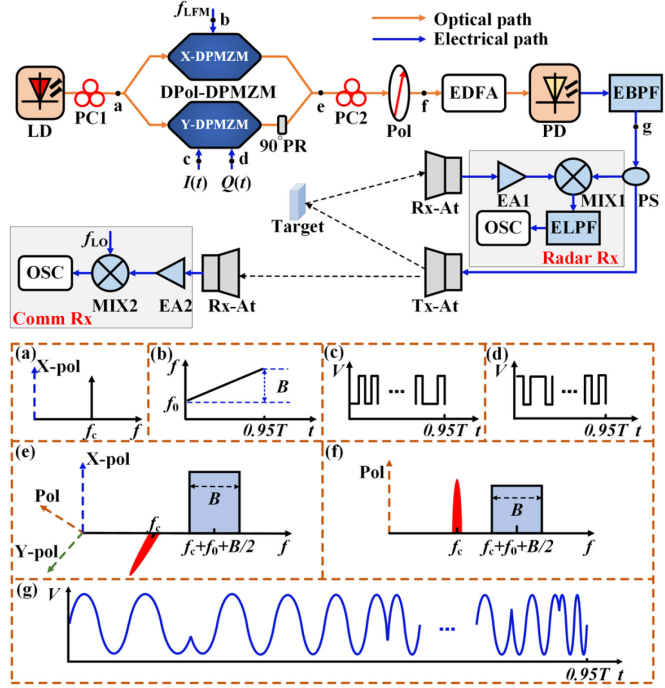

Fig. 1. Schematic diagram of the proposed joint communication-radar system. LD, laser diode; PC, polarization controller; DPol-DPMZM, dual-polarization dual-parallel Mach-Zehnder modulator; PR, polarization rotator; EDFA, erbium-doped fiber amplifier; Pol, polarizer; PD, photodetector; EBPF, electrical band-pass filter; PS, power splitter; ELPF, electrical low-pass filter; EA, electrical amplifier; MIX, mixer; OSC, oscilloscope; Tx-At, transmitting antenna; Rx-At, receiving antenna. (a)-(g) are schematic diagrams of signals at different positions in the system diagram.

The schematic diagram of the proposed joint communication-radar system is shown in Fig. 1. A continuous-wave (CW) light wave $E_{in}(t)$ with a center frequency of $f_c$ from a laser diode (LD) is sent to a DPol-DPMZM via a polarization controller (PC1). The optical signal from the LD can be expressed as

$$E_{in}(t) = \sqrt{2}E_{in}\exp(j2\pi f_c t), \quad (1)$$

where $\sqrt{2}E_{in}$ and $f_c$ are the amplitude and frequency of the optical carrier, respectively. The DPol-DPMZM consists of two DPMZMs and each DPMZM consists of two sub-MZMs and a main-MZM. An LFM signal is sent to the DPMZM in the X polarization (X-DPMZM) via a 90° electrical hybrid coupler. The instantaneous frequency of the LFM signal can be expressed as

$$f_{LFM}(t) = \begin{cases} f_0 + kt, & t \in (0, 0.95T] \\ 0, & t \in (0.95T, T] \end{cases}, \quad (2)$$

where $f_0$ is the initial frequency, $B$ is the bandwidth, and $k$ is the chirp rate of the LFM signal. The period and duty ratio of the LFM signal is $T$ and 95%, so $k=B/0.95T$. The time-frequency diagram of the LFM signal is shown in Fig. 1(b). The single-chirped LFM signals from the 90° electrical hybrid coupler can be expressed as

$$V_{RF1}(t) = V_{RF} \sin\left[2\pi\left(f_0 t + \frac{1}{2}kt^2\right)\right], \qquad (3)$$

$$V_{RF2}(t) = V_{RF} \cos\left[2\pi\left(f_0 t + \frac{1}{2}kt^2\right)\right], \qquad (4)$$

where $V_{RF}$ is the amplitude of the LFM signal. Thus, the optical signal from the X-DPMZM can be expressed as [25]

$$\begin{aligned}
E_{X-DPMZM}(t) &= E_{MZM1} + E_{MZM2} \exp\left(j\pi \frac{V_3}{V_\pi}\right) \\
&= \frac{\sqrt{2}}{4} E_{in}(t) \Bigg\{ \cos\left[\frac{\pi}{2V_\pi}(V_{RF1}(t) + V_1)\right] \\
&\quad \times \exp\left(j\pi \frac{V_1}{2V_\pi}\right) + \cos\left[\frac{\pi}{2V_\pi}(V_{RF2}(t) + V_2)\right] \\
&\quad \times \exp\left(j\pi \frac{V_2}{2V_\pi}\right) \exp\left(j\pi \frac{V_3}{V_\pi}\right) \Bigg\},
\end{aligned} \qquad (5)$$

where $E_{MZM1}$, $E_{MZM2}$ is the output optical signals of the sub-MZMs, $V_\pi$ is the half-wave voltage of the modulator, $V_1$, $V_2$, and $V_3$ denote the three biases applied to the sub-MZMs and the main-MZM, respectively. To implement CS-SSB modulation, both sub-MZMs are biased at the minimum transmission points and the main-MZM is biased at the quadrature transmission point, i.e., $V_1 = V_2 = -V_\pi$, $V_3 = V_\pi/2$. Under these circumstances, (5) can be rewritten as

$$\begin{aligned}
E_{X-DPMZM}(t) &= \frac{1}{2} E_{in} \Bigg\{ \cos\left[\frac{\pi}{2V_\pi} V_{RF} \cos(2\pi f_0 t + \pi k t^2) - \frac{\pi}{2}\right] \\
&\quad \times \exp(j 2\pi f_c t) \\
&\quad - j \cos\left[\frac{\pi}{2V_\pi} V_{RF} \sin(2\pi f_0 t + \pi k t^2) - \frac{\pi}{2}\right] \\
&\quad \times \exp(j 2\pi f_c t) \Bigg\} \\
&\approx \frac{1}{2} m_1 E_{in} \exp\left[j 2\pi\left(f_c t - f_0 t - \frac{1}{2}kt^2\right)\right],
\end{aligned} \qquad (6)$$

where $m_1 = \pi V_{RF}/2V_\pi$ is the modulation index. In the deduction of (6), small-signal modulation is used, so higher-order modulation sidebands are ignored. As can be seen from (6), only the -1st-order optical sideband of the LFM signal is generated after the X-DPMZM, with the optical spectrum shown in the X polarization in Fig. 1(e).

The DPMZM in the Y polarization (Y-DPMZM) is used to implement optical IQ modulation on the optical carrier [26]. It is assumed that the two data streams applied to the Y-DPMZM are $I(t)$ and $Q(t)$, which are shown in Fig. 1(c) and (d), respectively. The Y-DPMZM is biased as an IQ modulator, which indeed has the same bias condition as the CS-SSB modulation for RF signals. Therefore, the optical signal at the output of the Y-DPMZM can be expressed as

$$E_{Y-DPMZM}(t) = \frac{\sqrt{2}}{4} E_{in}(t) \left\{ \cos\left[\pi \frac{(I(t)-V_\pi)}{2V_\pi}\right] + \right.$$

$$\left. \cos\left[\pi \frac{(Q(t)-V_\pi)}{2V_\pi}\right] \exp\left(j\frac{\pi}{2}\right) \right\} \quad (7)$$

$$\approx \frac{1}{2} m_2 E_{in} \left[I(t) + jQ(t)\right] \exp(j2\pi f_c t)$$

$$= \frac{1}{2} m_2 E_{in} g(t) \exp\left[j2\pi f_c t + j\varphi(t)\right],$$

where $m_2 = \pi/2V_\pi$, $g(t) = \sqrt{I^2(t)+Q^2(t)}$, $\varphi(t) = \arctan\left[Q(t)/I(t)\right]$. In the deduction of (7), small modulation condition is also considered. It is shown that the data signals are IQ modulated onto the optical carrier, as shown in the Y polarization in Fig. 1(e).

At the output of the DPol-DPMZM, by adjusting PC2, the polarization state of the X-polarized light or the Y-polarized light from the DPol-DPMZM is oriented to have an angle of 45° to the main axis of the polarizer (Pol) to combine the optical signals in the two polarizations. In this case, the optical signals at the output of the Pol can be expressed as

$$E_1(t) = \frac{\sqrt{2}}{2} E_{X-DPMZM}(t) + \frac{\sqrt{2}}{2} E_{Y-DPMZM}(t)$$

$$= \frac{\sqrt{2}}{4} m_1 E_{in} \exp\left[j2\pi\left(f_c t - f_0 t - \frac{1}{2}kt^2\right)\right] \quad (8)$$

$$+ \frac{\sqrt{2}}{4} m_2 E_{in} g(t) \exp\left[j2\pi f_c t + j\varphi(t)\right].$$

According to (8), the optical signal from the Pol consists of an IQ-modulated optical carrier and an optical sideband carrying the LFM signal, which is shown in Fig. 1(f).

The optical signal from the Pol is amplified by an erbium-doped fiber amplifier (EDFA) and then sent to a PD, whose electrical output can be expressed as

$$i(t) = \eta |E_1(t)|^2$$

$$= \underbrace{\frac{1}{8}\eta m_1^2 G^2 E_{in}^2 + \frac{1}{8}\eta m_2^2 G^2 E_{in}^2 g^2(t)}_{low-frequency\ component} \quad (9)$$

$$+ \underbrace{\frac{1}{4}\eta m_1 m_2 G^2 E_{in}^2 g(t) \cos\left[2\pi\left(f_0 t + \frac{1}{2}kt^2\right) + \varphi(t)\right]}_{desired\ high-frequency\ joint\ communication-radar\ signal},$$

where $\eta$ is the responsivity of the PD, and $G$ is the gain of the EDFA. As can be seen, the photocurrent consists of two low-frequency components and a desired high-frequency component, which can be used as the joint communication-radar signal. The bit rate $R_b$ of $I(t)$ and $Q(t)$ is set to multiples of the repetition rate of the LFM signal so that the desired frequency component is an LFM signal sliced by an IQ data signal. The waveform of the IQ-data-sliced LFM signal in one period is schematically shown in Fig. 1(g). The IQ-data-sliced LFM signal of interest is filtered out by an electrical band-pass filter (EBPF) and then divided into two parts by a power splitter (PS). The IQ-data-sliced LFM signal from one output of the PS serves as the transmitted signal of the joint communication-radar system via a transmitting antenna. The transmitted joint communication-radar signal is further expressed as

$$V_{TX}(t) = \xi(t)\cos\left[2\pi\left(f_0 t + \frac{1}{2}kt^2\right) + \varphi(t)\right], \tag{10}$$

where $\xi(t) = 0.25R\eta m_1 m_2 G^2 E_{in}^2 g(t)$ and $R$ is the load resistance. The key difference between the joint communication-radar signal in (10) and a conventional IQ-modulated signal is that the single-frequency carrier of the conventional IQ modulation is replaced by an LFM carrier.

In the radar receiver, the electrical signal from the other output of the PS is connected to the local oscillator (LO) port of an electrical mixer (MIX1) for de-chirping the echo signals received by the radar receiving antenna and amplified by an electrical amplifier (EA1). The delay time of the radar echo signals can be expressed as

$$\Delta\tau = \frac{2L_0}{c}, \tag{11}$$

where $c$ is the velocity of light in a vacuum, and $L_0$ is the distance between the antenna and the target. Thus, the radar echo signal applied to MIX1 can be expressed as

$$V_{echo}(t) = \zeta(t)\cos\left\{2\pi\left[f_0(t-\Delta\tau) + \frac{1}{2}k(t-\Delta\tau)^2\right] + \varphi(t-\Delta\tau)\right\}, \tag{12}$$

where $\zeta(t) = 0.25 l_1 G_1 G_T G_{R1} R\eta m_1 m_2 G^2 E_{in}^2 g(t)$, $l_1$ is the loss of the echo signal in free space, $G_1$ is the gain of the EA1, and $G_T$ and $G_{R1}$ are the antenna gains of the transmitting and radar receiving antenna, respectively. The de-chirped signal from MIX1 is filtered by an electrical low-pass filter (ELPF), whose output can be written as

$$V_{ELPF}(t) = \frac{1}{2}\xi(t)\zeta(t)\cos\left\{2\pi f_0 \Delta\tau + \pi k\left[t^2 - (t-\Delta\tau)^2\right] + \varphi(t) - \varphi(t-\Delta\tau)\right\}. \tag{13}$$

Compared with the de-chirped signal of a conventional LFM signal, the de-chirped signal of the IQ-data-sliced LFM signal contains an additional phase term in (13). It should be noted that the phase terms have an impact on radar performance, which will be discussed in the discussion section. Without considering the additional phase term, the frequency of the de-chirped signal from (13) can be expressed as

$$f_d \approx k\Delta\tau, \tag{14}$$

so, the distance of the target $L_0$ can be calculated as

$$L_0 = \frac{c}{2k}f_d. \tag{15}$$

Besides, real-time ISAR imaging can also be implemented based on the proposed system. When the radar transmits a sequence of $N$ pulses and each echo signal has $M$ samples, the de-chirped signal can be rearranged as an $M \times N$ matrix to construct a two-dimension image [27], [28]. Theoretically, the range resolution and the cross-range resolution of the ISAR imaging [29] can be expressed as

$$R_L = \frac{c}{2B}, \tag{16}$$

$$R_C = \frac{\lambda}{2T_r\Omega}, \tag{17}$$

where $\lambda$ is the center wavelength of the transmitted signal, and $T_r$ and $\Omega$ are the integration time and the rotating speed of the target in a frame image.

In the communication receiver, the IQ-data-sliced LFM signal from the transmitting antenna is received by a communication receiving antenna, amplified by EA2, and then applied to the RF port of the MIX2. A single-tone signal is sent to the LO port of the MIX2 to down-convert the received signal. The IF signal from MIX2 can be expressed as

$$V_{MIX2}(t) = \chi \left\{ I(t)\cos\left[2\pi\left(f_0 t - f_{LO}t + \frac{1}{2}kt^2\right)\right] \right. \\ \left. - Q(t)\sin\left[2\pi\left(f_0 t - f_{LO}t + \frac{1}{2}kt^2\right)\right] \right\}, \quad (18)$$

where $\chi = 0.125 l_2 G_2 G_T G_{R2} R \eta m_1 m_2 G^2 E_{in}^2 E_{LO}$, $l_2$ is the loss of the IQ-data-sliced LFM signal in free space, $G_2$ is the gain of the EA2, $G_T$ and $G_{R2}$ are the antenna gains of the transmitting and communication receiving antenna, and $E_{LO}$ is the amplitude of the LO signal. Then, an IF-LFM signal without data modulation is used for coherent demodulation of the IQ-data-sliced LFM signal in the digital domain. The coherent demodulation process can be expressed as

$$I_1(t) = V_{MIX2}(t)\cos\left[2\pi\left(f_0 t - f_{LO}t + \frac{1}{2}kt^2\right)\right] \\ = \underbrace{\frac{1}{2}\chi I(t)}_{desired\ baseband\ component} + \underbrace{\frac{1}{2}\chi I(t)\cos\left[4\pi\left(f_0 t - f_{LO}t + \frac{1}{2}kt^2\right)\right]}_{high-frequency\ component} \\ - \underbrace{\frac{1}{2}\chi Q(t)\sin\left[4\pi\left(f_0 t - f_{LO}t + \frac{1}{2}kt^2\right)\right]}_{high-frequency\ component}, \quad (19)$$

$$Q_1(t) = V_{MIX2}(t)\sin\left[2\pi\left(f_0 t - f_{LO}t + \frac{1}{2}kt^2\right)\right] \\ = \underbrace{-\frac{1}{2}\chi Q(t)}_{desired\ baseband\ component} + \underbrace{\frac{1}{2}\chi I(t)\sin\left[4\pi\left(f_0 t - f_{LO}t + \frac{1}{2}kt^2\right)\right]}_{high-frequency\ component} \\ + \underbrace{\frac{1}{2}\chi Q(t)\cos\left[4\pi\left(f_0 t - f_{LO}t + \frac{1}{2}kt^2\right)\right]}_{high-frequency\ component}. \quad (20)$$

In (19) and (20), besides two high-frequency components, both output signals contain the desired baseband component, which can be filtered out by a digital low-pass filter to recover the baseband data streams for wireless communications.

## 4. Experimental results and discussion

*3.1. Experimental Setup*

An experiment based on the setup shown in Fig. 1 is carried out to verify the feasibility of the proposed system. The CW light wave from the LD (HLT-ITLA-M-C-1-FA) with a center wavelength of 1550.87 nm and optical power of 13 dBm is injected into the DPol-DPMZM (Fujitsu FTM7977) via PC1. The LFM signal generated by an arbitrary waveform generator (AWG, Keysight 8195A) is sent to the X-DPMZM in the DPol-DPMZM via the 90° electrical hybrid coupler (Narda 4065 7.5-16 GHz) to realize the CS-SSB modulation. Two pseudo-random bit sequences, i.e., the IQ data streams, with a bit rate of

105.26 Mbit/s are also generated from the AWG, which are applied to the RF ports of the Y-DPMZM in the DPol-DPMZM, to implement IQ modulation on the optical carrier. By adjusting PC2, the two orthogonally polarized lights from the DPol-DPMZM are combined at the Pol and then sent to the EDFA (Amonics AEDFA-PA-35-B-FA) to for power compensation. The amplified optical signal from EDFA is injected into the (Nortel PP-10G) to generate the QPSK-sliced LFM signal. An EBPF (7.91-11.1 GHz) is utilized following the PD to filter out the out-of-band interference signals. Then, the QPSK-sliced LFM signals are divided into two parts by the PS (Narda 4456, 2-18 GHz). One output of the PS is fed directly to a transmitting antenna (GHA080180-SMF-14, 8-18 GHz) as the transmitted signal and the other output of the PS is connected to the LO port of MIX1 (Miteq m30) as a reference signal. At the radar receiver, the echo signal reflected from the target is collected by a radar receiving antenna (GHA080180-SMF-14, 8-18 GHz) and then amplified by EA1 (ALM/145-5023-293, 5.85-14.50 GHz). The amplified echo signal is sent to the RF port of MIX1. The de-chirped signal from MIX1 is filtered by an ELPF (DC-1.4 GHz) and captured by an oscilloscope (OSC, R&S RTO2032), which is further post-processed to implement radar functions. In the communication receiver, the transmitted signal is collected by a communication receiving antenna (GHA080180-SMF-14, 8-18 GHz) and then amplified by EA2 (ALM/145-5023-293, 5.85-14.50 GHz). The amplified QPSK-sliced LFM signal is down-converted to the IF band by an 8-GHz single-tone signal generated from the AWG. The down-converted QPSK-sliced LFM signal is captured by the OSC and then digitally demodulated.

*3.2. Joint communication-radar signal generation*

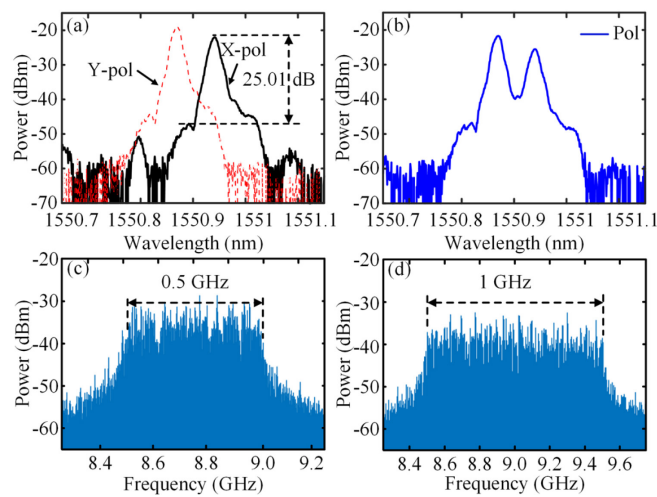

Fig. 2. (a) Optical spectra of the RF-modulated and IQ-modulated optical carrier, (b) optical spectrum of the optical signal from the Pol, electrical spectra of the QPSK-sliced LFM signal centered at (c) 8.75 GHz and (d) 9.0 GHz.

First, an LFM signal from 8.5 to 9.5 GHz is generated by the AWG. The width and period of the LFM signals are 3.8 μs and 4 μs, respectively. The optical spectrum of the optical signal from the X polarization of the DPol-DPMZM after CS-SSB modulation is shown in black solid line in Fig. 2(a), in which the optical carrier and the +1st-order optical sideband are both deeply suppressed by more than 25 dB. The optical carrier in the Y polarization of the DPol-DPMZM after IQ modulation is shown in the red dashed line in Fig. 2(a). Due to the limited resolution of the optical spectrum analyzer (OSA, Ando 6317B), the data modulation on the optical carrier with a speed of 105.26 Mbit/s and the linearly sweeping -1st-order optical sideband cannot be observed from the spectrum.

After combination at the Pol, the optical spectrum of the combined optical signal from the Pol is shown in Fig. 2(b), which consists of an IQ-modulated optical carrier and a -1st-order optical sideband carrying the LFM signal. The combined optical signal is detected in the PD after amplification in the EDFA. The spectrum of the photocurrent from the PD is captured by an electrical spectrum analyzer (ESA, R&S FSP-40). It is seen in Fig. 2(c) that the QPSK-sliced LFM signal with a bandwidth of 0.5 GHz (from 8.5 to 9 GHz) is generated. To verify the tunability of the system, the QPSK-sliced LFM signal with a bandwidth of 1 GHz (from 8.5 to 9.5 GHz) is also generated, as shown in Fig. 2(d).

### 3.3. Radar function

The photograph of the experimental setup for radar function is shown in Fig. 3(a). To achieve distance measurement, two cuboids are used as static targets as shown in Fig. 3(b), where *d* is the distance between the two cuboids and *D* is the distance between the antennas and the closer target. First, the two cuboids are placed at a distance of 0.96 and 1.29 m along the radar line of sight, i.e., *d=0.33 m* and *D=0.96 m*. The 105.26-Mbit/s QPSK-sliced LFM signal with a bandwidth of 0.5 GHz (from 8.5 to 9 GHz) is transmitted via the transmitting antenna. Note that, in this paper, this bandwidth means the bandwidth of the LFM carrier for radar function and the data rate represents the communication data rate, i.e., $I(t)$ or $Q(t)$ for communication function. The de-chirped signal from MIX1 is monitored by the OSC with a sampling rate of 20 MSa/s. After Fourier transforms, the spectrum of the de-chirped signal in one pulse period (4 μs) is plotted in Fig. 3(c). Two individual peaks can be observed, which are located at 0.82 and 1.17 MHz, respectively. According to (15), the distances of the two targets are 0.93 and 1.33 m, respectively, which are close to actual values of 0.96 and 1.29 m. Secondly, two cuboids are placed at a distance of 0.96 and 1.13 m, i.e., *d=0.17 m* and *D=0.96 m*. In this experiment, the bandwidth of the 105.26-Mbit/s QPSK-sliced LFM signal is 1 GHz (from 8.5 to 9.5 GHz). The spectrum of the de-chirped signal after Fourier transform is plotted in Fig. 3(d). Two individual peaks can also be observed, which are located at 1.69 and 2.02 MHz, respectively. The distances of the two targets according to (15) are 0.96 and 1.15 m, respectively, which are close to actual values of 0.96 and 1.13 m.

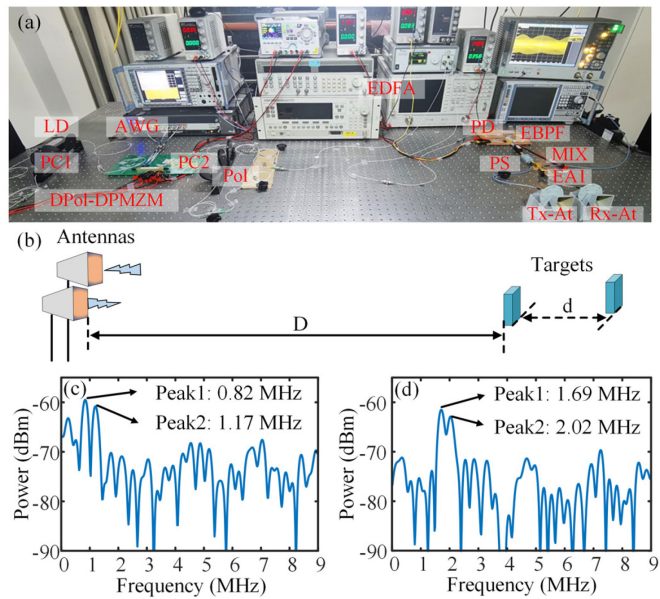

Fig. 3. (a) The schematic diagram of the distance measurement, (b) the photograph of the experimental setup for radar function, (c) electrical spectrum of the de-chirped signal when *d* is 0.33 m, (d) electrical spectrum of the de-chirped signal when *d* is 0.17 m.

With the proposed system, not only can the distance of the targets be measured, but also ISAR imaging can be implemented. The imaging targets are placed in different positions on a turntable as shown in Fig. 4(a), where $L$ is the distance between the antenna pair and the center of the turntable. In this experiment, a 105.26-Mbit/s QPSK-sliced LFM signal with the bandwidth of 1 GHz (from 8.5 to 9.5 GHz) is transmitted. One cuboid and two cylinders as the targets are placed in different positions on the turntable, which is shown in Fig. 4(b). Along the radar line of sight, the center of the turntable is about 1.23 m away from the antenna pair. The de-chirped signal is sampled with a sampling rate of 20 MSa/s and a sampling time of 2 s. According to (16) and (17), the theoretical range and cross-range resolutions are 14.99 and 3.25 cm, respectively. Fig. 4(c) shows the imaging result of the three targets. As can be seen, three targets can be clearly distinguished.

Then, the bandwidth of the signal and the distance from the turntable center to the antenna pair are changed to further verify the feasibility of the system. In this experiment, the transmitted signal is a 105.26-Mbit/s QPSK-sliced LFM signal with a bandwidth of 0.5 GHz (from 8.5 to 9 GHz), and two cuboids as the targets are placed in different positions on the turntable, which is shown in Fig. 4(d). Along the radar line of sight, the center of the turntable is about 1.43 m away from the antenna pair and the de-chirped signal is sampled with a sampling rate of 20 MSa/s and a sampling time of 1 s. According to (16) and (17), the theoretical range resolution and the cross-range resolution are 29.98 and 6.70 cm, respectively. Fig. 4(e) shows the imaging result of the two targets. As can be seen, two targets can be distinguished. The real-time imaging capability of the radar system is verified.

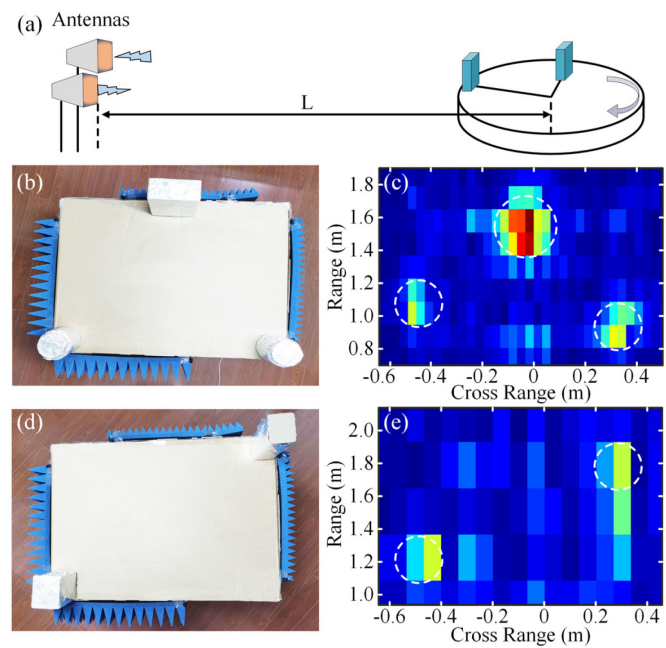

Fig. 4. (a) The schematic diagram of the ISAR imaging, (b) the photograph of the three targets under test, (c) the imaging result of the three targets, (d) the photograph of the two targets under test, (e) the imaging result of the two targets.

## 3.4. Communication function

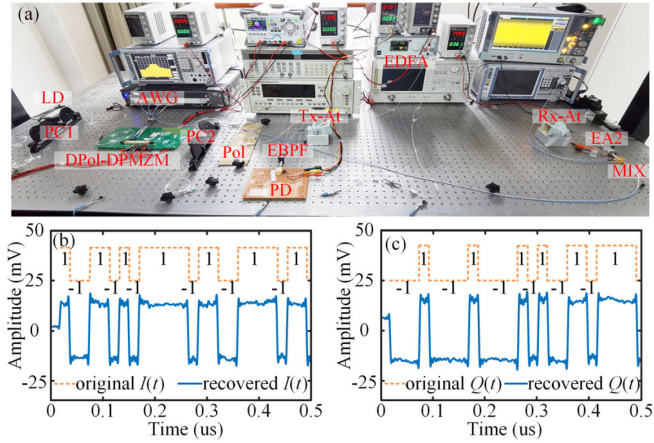

Fig. 5. (a) The photograph of the experimental setup for communication function, (b) the baseband signal recovered from the in-phase branch compared with $I(t)$, (c) the baseband signal recovered from the quadrature branch compared with $Q(t)$.

The communication function of the joint communication-radar system is further demonstrated. The photograph of the experimental setup for communication function is shown in Fig. 5(a). First, the communication function is verified by a wired connection via replacing the two antennas in Fig. 5(a) using a 1.5-m RF cable. In this experiment, a 105.26-Mbit/s QPSK-sliced LFM signal with a bandwidth of 0.5 GHz (from 8.5 to 9 GHz) is transmitted, which is down-converted to 0.5 to 1 GHz using an 8-GHz LO signal. The down-converted QPSK-sliced IF-LFM signal is sampled by the OSC at a sampling rate of 10 GSa/s, which is digitally demodulated by post-processing.

The baseband signals recovered from the in-phase branch and the quadrature branch are shown in the blue solid line in Fig. 5(b) and (c). To show it more clearly, the original bits from 0 to 0.5 μs in the two branches are also given. Compared with the original $I(t)$ and $Q(t)$ as shown in orange dashed line in Fig. 5(b) and (c), it can be seen that the original data streams are well recovered from the received signal, which proves the feasibility of the communication function of the QPSK-sliced LFM signal.

| Bit rates / Bandwidths | 105.26 Mbit/s | 210.52 Mbit/s |
|---|---|---|
| 0.5 GHz | EVM:12.27% | EVM:12.02% |
| 1.0 GHz | EVM:15.14% | EVM:15.19% |

Fig. 6. EVMs and constellation diagrams of the QPSK-sliced LFM signal at different bandwidths and different bit rates.

The error vector magnitude (EVM) of the received signal is further used to show the performance of the communication function. When the received signal power is -16 dBm, the constellation diagrams and EVMs of the received signals at different bandwidths and bit rates are shown in Fig. 6. As can be seen, when the data rate of the IQ streams changes from 105.26 to 210.52 Mbit/s, the EVM does not have obvious variation under the same LFM bandwidth. However, under the same communication data rate, the EVM gets worse when the LFM bandwidth is increased. The reason for this phenomenon is that as the bandwidth of the QPSK-sliced LFM signal increases, the flatness of the QPSK-sliced LFM signal becomes worse, resulting in a worse amplitude flatness of the demodulated signals.

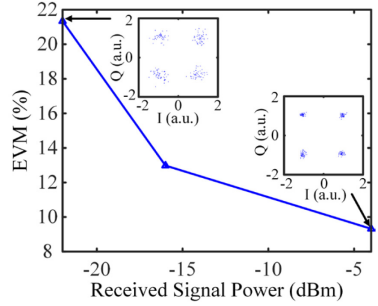

Fig. 7. Measured EVM versus the received signal power. The insets are the constellation diagrams when received signal power is -22 dBm and -4 dBm.

Then the EVM versus the received signal power is evaluated. In this experiment, different transmission conditions are used to obtain different received signal power. The received signal power is -4, -16 and -22 dBm for 1.5-m wired transmission, 0.35-m wireless transmission, and 0.75-m wireless transmission, respectively. The 105.26-Mbit/s QPSK-sliced LFM signal with a bandwidth of 0.5 GHz (from 8.5 to 9 GHz) are used and the corresponding results are shown in Fig. 7. The insets in Fig. 7 show the constellation diagrams when the received signal power is -22 and -4 dBm. It is shown that the EVM is 9.32%, 12.98%, and 21.34% when the received signal power is -4, -16, and -22 dBm, respectively.

*3.5. Discussions*

Compared with the de-chirped signal of a conventional LFM signal, the de-chirped signal of the IQ-data-sliced LFM signal contains an additional phase term in (13), i.e., $\varphi(t)-\varphi(t-\Delta\tau)$, which is related to the distance of the radar target. As is well known, ISAR imaging takes full use of the frequency and phase of the de-chirped signal, so the additional phase term has a corresponding impact on ISAR imaging. When the delay time $\Delta\tau$ is much smaller than the symbol length of the IQ-data-sliced LFM signal, i.e., $\varphi(t)-\varphi(t-\Delta\tau)\approx 0$ establishes, the impact of the additional phase term can be ignored. When the delay gradually increases to be comparable to or even much larger than the symbol length, phase disturbances will be introduced in the imaging algorithm, resulting in the degradation of imaging quality until it cannot image.

Therefore, the distance of the imaging targets is limited by the symbol length. In the experiment, for the joint communication-radar signal with the symbol length of around 20 ns, the targets less than 1.5 m away from the antenna pair are successfully imaged. As a comparison, a joint communication-radar signal with a symbol length of 10 ns is used to image the same targets and the experimental results were not satisfactory, which is caused by the additional phase term that cannot be ignored. Therefore, in real-world applications, there is a trade-off between the target distance in ISAR imaging and the bit rate in communication.

In comparison, when only the distance measurement is required in the radar detection, the additional phase term has little influence on the distance measurement results because distance measurement only uses the frequency information of the de-chirped signal.

Therefore, when only distance measurement and communication functions are required in the joint communication-radar system, the limitations of the distance of targets and the communication data rate are greatly relaxed. However, when the ISAR imaging function is enabled, the communication data rate should be limited; when the ISAR imaging function is disabled, the communication data rate can be enhanced.

## 4. Conclusion

In summary, a joint communication-radar system based on a QPSK-sliced LFM signal is proposed and experimentally demonstrated for both radar and communication functions. The key significance of the work is that the QPSK-sliced LFM signal proposed in this work can not only achieve high spectral efficiency data transmission using the quaternary phase coding, but also easily realize ISAR imaging and radar range detection by using the LFM signal at the same time. An experiment is carried out. Radar range detection with an error of less than 4 cm, ISAR imaging with a resolution of 14.99 cm × 3.25 cm, and communication with a data rate of 105.26 Mbit/s are successfully verified. The system and signal format proposed in this paper will provide a feasible solution for the future joint communication-radar system.

**Funding**

National Natural Science Foundation of China (NSFC) (61971193); Natural Science Foundation of Shanghai (20ZR1416100); Open Fund of State Key Laboratory of Advanced Optical Communication Systems and Networks, Peking University, China (2020GZKF005); Science and Technology Commission of Shanghai Municipality (18DZ2270800).

**Conflicts of interest**

The authors declare no conflicts of interest.